# Solar activity and Svalbard temperatures

Jan-Erik Solheim,[1] Kjell Stordahl[2], Ole Humlum[3,4]

1) *Institute of Theoretical Astrophysics, University of Oslo, Oslo, Norway,*
*j.e.solheim@astro.uio.no*
2) *Telenor Norway, Finance, Fornebu, Oslo, Norway,*
*kjell.stordahl@telenor.com*
3) *Department of Geosciences, University of Oslo, Oslo, Norway,*
*Ole.Humlum@geo.uio.no*
4) *Department of Geology, University Centre in Svalbard (UNIS), Svalbard.*

**Abstract**

The long temperature series at Svalbard (Longyearbyen) show large variations, and a positive trend since its start in 1912. During this period solar activity has increased, as indicated by shorter solar cycles. The temperature at Svalbard is negatively correlated with the length of the solar cycle. The strongest negative correlation is found with lags 10-12 years.

The relations between the length of a solar cycle and the mean temperature in the following cycle, is used to model Svalbard annual mean temperature, and seasonal temperature variations. Residuals from the annual and winter models show no autocorrelations on the 5 per cent level, which indicates that no additional parameters are needed to explain the temperature variations with 95 per cent significance. These models show that 60 per cent of the annual and winter temperature variations are explained by solar activity. For the spring, summer and fall temperatures autocorrelations in the residuals exists, and additional variables may contribute to the variations.

These models can be applied as forecasting models. We predict an annual mean temperature decrease for Svalbard of 3.5±2 °C from solar cycle 23 to solar cycle 24 (2009-20) and a decrease in the winter temperature of ≈6 °C.

1. **Introduction**

The question of a possible connection between solar activity and the Earth's climate has received considerable attention the last 200 years [1,2]. It may therefore be of interest to investigate if part of the temperature increase and large variations in Arctic temperatures may be attributed to solar influence.

One of the longest Arctic temperature series is from Svalbard. It started in 1912 and is discussed and analysed by Humlum, Solheim and Stordahl [3] (called HSS12 in the following). Their analysis of the temperature record identified a linear trend of 0.023 °C yr$^{-1}$, in addition to cyclic variations. The strongest cyclic variations have periods 62-68, 26, and 15-17 years. HSS12 also finds lower amplitude variations with periods 11-12 years, which may be related to the solar activity cycle.

Solar activity can be described by the following proxies described by Gray et al. [2] (their figures 1 and 2): The sunspot number $R$; the 10.7 cm solar radio flux, $F_{10.7}$; the chromospheric Mg II line; the open solar flux near the Earth $F_s$; galactic cosmic ray neutron counts; the total solar irradiance (TSI); variations in the geomagnetic field (the *aa* index); counts of low latitude aurora; and the radioactive $^{10}$Be isotope. Most attention has been paid to the number of sunspots, which varies with a period of 9-13 years. Also the length of the sunspot cycle is known to vary with the solar activity, in the sense that high activity is related to short cycles and low activity to long cycles. The length is shown to vary in a systematic way in a cycle of length of 80-90 years, named after Gleissberg [4].

The length of a solar cycle can be determined from the appearance of the first spot in a cycle at high solar latitude, to the disappearance of the last spot in the cycle near solar equator. However, before the last spot in a cycle disappears, the first spot in the next cycle appears at high latitude, and there is normally a two years overlap [5]. The time of minimum is defined as the central time of overlap between the two cycles [5], and the length of a cycle can be measured between successive minima or maxima.

It was for a long time thought that the appearance of a solar cycle was a random event, i.e. each cycle's length and amplitude were independent of the previous. However, Dicke [6] showed already in 1978, that an internal chronometer has to exist inside the Sun, which after a number of short cycles, reset the cycle length so the average length of 11.2 years is kept.

Comparing sunspot numbers with the Northern Hemisphere land temperature anomaly, Friis-Christensen & Lassen [7] found a better correlation between the solar cycle length (SCL) and the temperature anomaly than with the number of sunspots, in the sense that shorter sunspot cycles indicated higher temperatures. They used a smoothed value for SCL over 5 solar cycles. This resulted in good correlations when the SCL on average decreased, indicating that solar activity was the dominant contributor to the NH-land temperature increase in their period of investigation (1860-1990).

However, at the turn of the millennium, inconsistency with this relation was found [8,9,10] because of the sudden lengthening of the solar cycle (see figure 1) while the NH-land temperature anomaly remained positive.

Solanki and Krivova [11] showed by cross correlations between time series of solar cycle length and sunspot number maxima ($R_{max}$), that the length precedes the amplitude, in the sense that short cycles preceded strong cycles. A test of the back time showed that a lag of one period of 11 years or one solar cycle, gave highest correlation after SC10. Before SC10 a lag of 3 cycles dominated. They interpreted this as the solar dynamo has a memory of the length of previous cycles, and suggested an empirical model based on lag one and three cycles for predicting the amplitude of the following cycle.

.

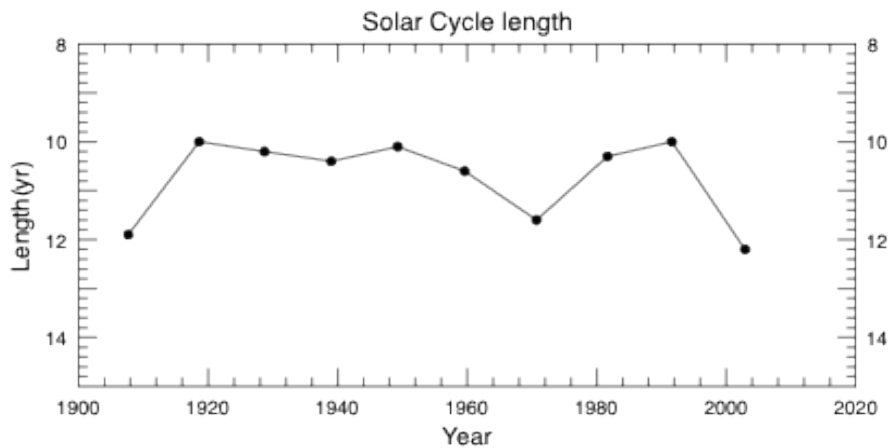

*Figure 1. Length of solar cycles (inverted scale) since 1900. The black dots are the mid-time for each cycle*

Butler [12] proposed that a well observed temperature series as the one at Armagh Observatory might be a better indicator for NH temperature anomaly than the HadCRUT3 average, which has a variable number of stations and includes large cities with urban heating. He found good correlations between the temperature measured at Armagh Observatory in the period 1844-1992 and the number of sunspots smoothed with the 1-2-2-2-1 filter. He concluded that solar activity, or something closely related to it, had dominating influence of the lower atmosphere temperature at Armagh in this period. Butler and Johnston [13], studied the same data and noticed a delay of about one solar cycle (10-12 yrs) between the shortest solar cycles and temperature peaks at Armagh. Archibald [14] proposed that this delay could be used to predict a cooling during the present solar cycle 24, which follows the longest in a century solar cycle 23, for certain locations in Europe and the East Coast of the USA where he found correlations between SCL and the temperature in the following cycle.

A systematic study by Solheim, Stordahl and Humlum [15] (called SSH11 in the following) of the correlation between SCL and temperature lags in 11-years intervals, for 16 data sets, revealed that the strongest correlation took place 10-12 years after the mid-time of a solar cycle, for most of the locations included. In this study the temperature series from Svalbard (Longyearbyen) was included, and a relation between the previous sunspot cycle length (PSCL) and the temperature in the following cycle was determined. This relation was used to predict that the yearly average temperature, which was -4.2 °C in sunspot cycle (SC) 23, was estimated to decrease to -7.8 °C in SC24, with a 95% confidence interval of -6.0 to -9.6 °C [15]. SSH11[15] found that stations in the North Atlantic (Torshavn, Akureyri and Svalbard), had the highest correlations between the length of a solar cycle and the temperature in the next cycle, showing correlation coefficient *r* from 0.79 to 0.86

In figure 2, we show the correlation *r* between the solar cycle length and an 11-year running mean temperature, with zero to 13 years lag for the Svalbard temperature series. The correlation is always negative and has a maximum absolute value between 10 and 13 years lag. This indicates that a lag of one solar cycle may give the highest correlation.

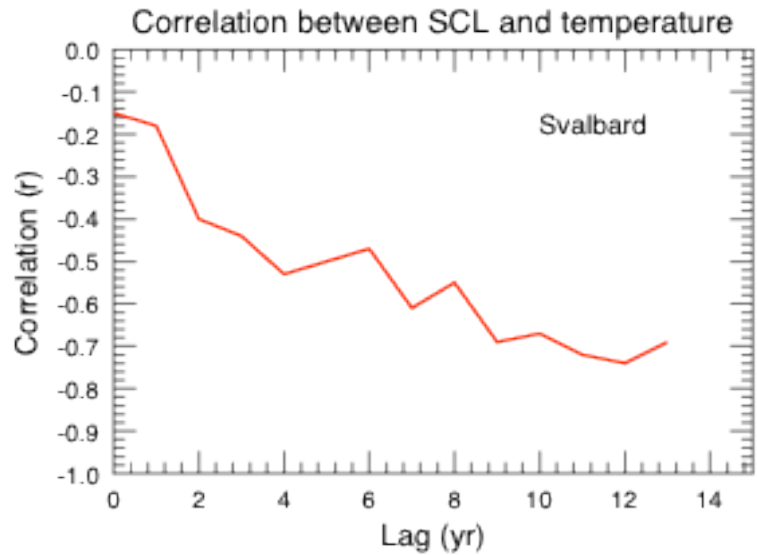

*Figure 2. Correlation r between the length of a solar cycle and 11-years running mean of the Svalbard (Longyearbyen) temperature record, with different lags.*

In the following we will discuss the Svalbard temperature series in more detail, and investigate how this correlation is related to seasons. We will also investigate how other variables may contribute to temperature variations at Svalbard.

**2. Svalbard temperature series and solar cycle relations**

*2.1 The Svalbard temperature record.*
The modern official Svalbard meteorological station is located near the main settlement in Svalbard, Longyearbyen (78° 13'N, 15° 33'E, about 2000 inhabitants), in central Spitsbergen. The station is located at the Svalbard Airport (24 m asl.), about 3 km NW of Longyearbyen, near the shore of the large fjord Isfjorden. Monthly temperature data were obtained from the *eKlima* portal run by the Norwegian Meteorological Institute. MAAT and seasonal temperature (DJF, MAM, JJA and SON) values 1912-2011 were calculated from this.

The Svalbard meteorological record is a composite record, representing homogenized observations originally made at 4 different stations, located along the shore of the fjord Isfjord, extending from the west coast to the interior of the main island Spitsbergen. A survey of meteorological statistics for the Norwegian Arctic is described by Førland et al. [16], and the individual Spitsbergen meteorological stations are described by Hanssen-Bauer et al. [17]. The Standard Normal Homogeneity Test [18,19] was applied on the series, and the results were validated by a study of the stations history by Nordli et al. [20]. The absence of visible irregularities in the record itself (Figure 3) as well as in wavelet diagrams [3], corresponding to the timing of known station changes, testifies to the quality of the homogenization carried out.

Comparing the Svalbard MAAT record with average Arctic temperature development since 1912 [21,22] there are both similarities and differences. The Arctic temperature increase around 1920 lasting to about 1940 is recognised in the Svalbard record, although this increase apparently began about 5 years before 1920. The general Arctic temperature decrease from about 1940 lasting to about 1970 is clearly visible in the Svalbard record also. Finally, the general Arctic temperature increase since 1980, until now, is also expressed by the Svalbard record, although with an apparent delay of 5-10 years.

*2.2 Svalbard temperatures in solar cycles.*

The Svalbard temperature series starts in 1912. The starting dates (in decimal years) and the length of solar cycles after 1900 are given in table 1 which is obtained from the National Geophysical Data Center (NGDC): ftp://ftp.ngdc.gov/STP/SOLAR_DATA/SUNSPOT_NUMBERS/INTERNATIONAL/maxmin/MAXMIN.

Solar cycle 14 (SC14) began in 1901 and was nearly finished when the Svalbard temperature observations started. Our analysis therefore starts with SC15 which began in 1914. Table 1 gives the mean temperature in each solar cycle, based on yearly temperatures in the range of years given in column 5.

TABLE 1: Svalbard: Mean temperatures in sunspot cycles

| Cycle no. | Mimimum yr | Rmax no | Lenght yr | Years temp. | year MAAT °C | Winter DJF °C | Spring MAM °C | Summer JJA °C | Fall SON °C |
|---|---|---|---|---|---|---|---|---|---|
| 14 | 1901.7 | 64.2 | 11.9 | | | | | | |
| 15 | 1913.6 | 105.4 | 10.0 | 1914-23 | -7.80 | -17.01 | -12.70 | 4.32 | -5.90 |
| 16 | 1923.6 | 78.1 | 10.2 | 1924-33 | -5.89 | -11.70 | -11.51 | 4.40 | -4.77 |
| 17 | 1933.8 | 119.2 | 10.4 | 1924-43 | -5.32 | -11.61 | -10.44 | 4.39 | -3.53 |
| 18 | 1944.2 | 151.8 | 10.1 | 1944-53 | -5.86 | -13.06 | -10.47 | 4.19 | -4.30 |
| 19 | 1954.3 | 201.3 | 10.6 | 1954-64 | -5.77 | -13.73 | -9.72 | 4.07 | -3.62 |
| 20 | 1964.9 | 110.6 | 11.6 | 1965-76 | -6.70 | -14.51 | -10.72 | 4.20 | -5.73 |
| 21 | 1976.5 | 164.5 | 10.3 | 1977-86 | -6.45 | -14.70 | -10.53 | 4.29 | -4.78 |
| 22 | 1986.8 | 158.5 | 10.0 | 1987-96 | -5.96 | -14.24 | -8.85 | 4.61 | -5.22 |
| 23 | 1996.9 | 120.8 | 12.2 | 1997-2008 | -4.20 | -11.20 | -8.33 | 5.45 | -3.03 |
| 24 | 2008.9 | | | | | | | | |

We have also calculated mean temperatures for four seasons as given in the table. Figure 3 shows the yearly temperatures and the solar cycle mean temperatures with standard errors ($\sigma_i$). Weighted linear least square fits to the solar cycle mean temperature values on the form $y = \beta x + \alpha$ gave the trends $\beta_1$ shown in figure 3 and in table 2. As weight for the observation *i* we used:

$w_i = N(1/\sigma_i)/\Sigma_i(1/\sigma_i)$ , for *i= 1..N*, where *N*=9 is the number of cycles.

The calculated the linear trend is given by:

$\beta = (\Sigma_i(x_i-x_{mean})(y_i-y_{mean})w_i)/(\Sigma_i(x_i-x_{mean})^2 w_i)$, where $\sigma_i$ is the standard deviation of the mean value of the temperature in cycle *i*.

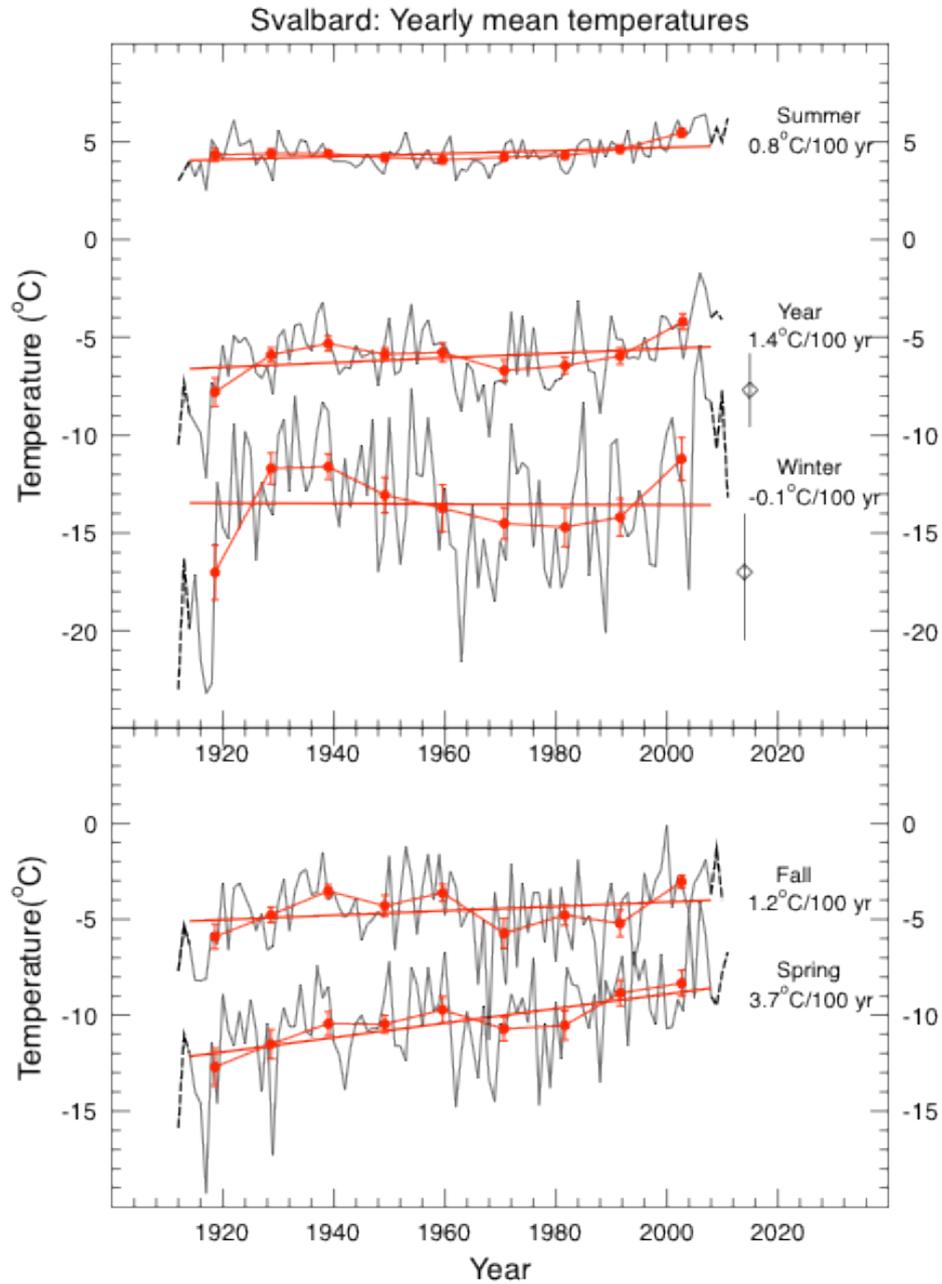

*Figure 3. The mean yearly (MAAT) temperatures at Svalbard (Spitzbergen) – for the year and the four seasons. The red circles are the mean temperatures in sunspot cycles with standard error bars, also used for the correlation analysis. The red thick lines are weighted linear least square fits to the solar cycle mean temperatures, with the trends for the period 1914-2008 ($\beta_1$) indicated. The black dashed curves are data before 1914 – and after 2008, corresponding to incomplete solar cycles, and not used in the averaging. Forecasts for SC24 temperatures based on length of SC23 are given with 95% confidence intervals (diamonds with bars) for the year and winter temperatures.*

The number of observations in the regression analysis is 9. Usually the degree of freedom is the number of observations minus the number of parameters in the model. However, analysis of the sunspot length effect on different lags of delayed temperature has been performed on beforehand, which resulted in a model where the previous sunspot cycle period explains the temperature in the next sunspot cycle. Hence, the regression model is reduced with one additional degree of freedom which results in 9 – 2 – 1 = 6 degrees of freedom.

The lengths of solar cycles since 1900 are shown with an inverted scale in figure 1 and can be compared with the Svalbard temperatures in figure 3. We recognize some qualitative similarities: The SCL shortened 1910-20, while the average temperature, and in particular the winter temperature, increased until about 1935. When the SCL became longer around 1970 a temperature minimum appeared a few years later. The short period SC22 which ended early in 1996 was followed by a temperature maximum around 2005.

Correlating the 11-years averaged Svalbard MAAT with the lengths of the solar cycles, shows that the correlation ($r$) increases in absolute value from 0.15 to 0.74 with increasing lags from 0 to 12 years (figure 2). The correlation increases to $r=0.79$ for the Svalbard yearly average temperature when the length of the previous solar cycle (PSCL) is correlated with the temperature in the next cycle, i.e. observed solar cycles are used instead of 11 years intervals [3].

*This motivates our choice of comparing the temperature in one solar cycle with the length of the previous cycle.*

*2.3 Correlations between the length of a solar cycle and the temperature in the next cycle.*

Linear least square fits between the length of a solar cycle and the average temperature in the next cycle (weighted with $w_i$), were done for the yearly and seasonal Svalbard mean temperature series (table 1). The resulting trends ($β_{PSCL}$) are shown in table 2, and the linear fits obtained are shown in figure 4.

Table 2 shows results of the model fitting for winter, spring, summer, fall and the whole year temperatures for the following models:
- Temperatures explained as a function of time (secular trend) (Model 1)
- Temperatures explained by the previous solar cycle (Model PSCL)

**TABLE 2.**        **Statistics on Svalbard temperature series**

|  | Model 1 | | Model PSCL | | Bootstrap (1000 samples) | |
| --- | --- | --- | --- | --- | --- | --- |
|  | $β_1$ °C(100 yr)$^{-1}$ | $r_1$ | $β_{PSCL}$ °Cyr$^{-1}$ | $r_{PSCL}$ | $r'_{PSCL}$ | 95% confidence limit |
| Year | 1.4±1.1 | 0.43 | -1.05±0.35 | 0.75 | 0.79 | 0.54:0.96 |
| Winter | -0.1±2.4 | 0.02 | -2.25±0.59 | 0.82 | 0.81 | 0.52:0.97 |
| Spring | 3.9±0.9 | 0.85 | -1.09±0.59 | 0.57 | 0.57 | 0.04:0.93 |
| Summer | 0.8±0.4 | 0.54 | -0.17±0.22 | 0.28 | 0.33 | 0.02:0.70 |
| Fall | 1.1±1.2 | 0.33 | -0.98±0.44 | 0.65 | 0.60 | 0.15:0.94 |

Analytical correlations coefficients ($r$) are calculated for Model 1 and Model PSCL, also shown in table 2. There is no analytical expression to estimate the error in the correlation coefficient $r$. We have therefore determined a correlation coefficient $r'$ by the so-called "bootstrap method." This is done by 1000 regression calculations on the *N* observations, by drawing sets of *N* observations (pairs of $x_i, y_i$) from the sample, and after each drawing return the observation to the sample. A regression coefficient is calculated for each new set, and the distribution of these regression coefficients are then analyzed. The mean value is the estimated non-parametric regression constant *r'*, and the 95% confidence interval is defined as the interval spanning from the 2.5th to the 97.5th percentile of the re-sampled *r*-values (*r'*).

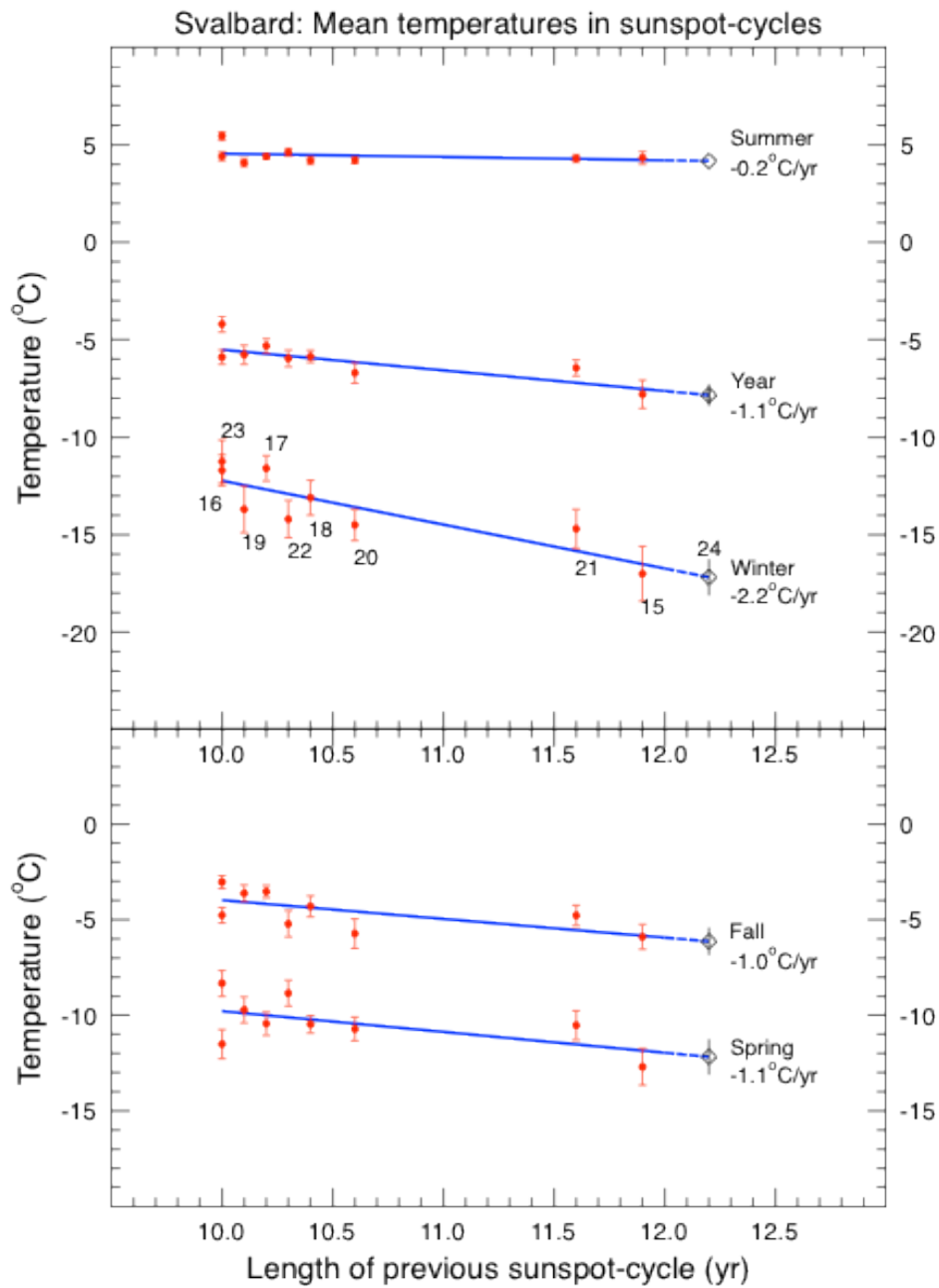

Figure 4. The average temperature during a solar cycle as a function of the length of the previous cycle for Svalbard yearly and seasonal mean temperatures 1914-2008. Trends ($\beta_{PSCL}$) are shown for each line. Solar cycle numbers are given for the winter series.

Comparing *r* and *r'*, we find they are on average are 6% and maximum 18% different. The coefficient of determination $r^2_{PSCL}$ is a measure of the contribution of the PSCL model to the temperature variations. For the Svalbard year and the winter mean temperatures $r^2_{PSCL}$= 0.62±0.35 within 95% confidence interval, which means that the Sun may contribute more than half to the temperature variations at Svalbard.

2.*4 Examination of correlations in the residuals*
The model fitting is not complete without examining the residuals. We have performed a Durbin-Watson (DW) statistical test [23,24,25] for serial correlation in the residuals - investigating if a positive error for one observation increases the chances of a positive error for another observation. The result is given in table 3. For two of the series, the yearly average and the winter temperatures, we find no autocorrelations in the residuals, which means that the model can be accepted by using the traditional statistical tests and confidence limits estimation without reduction of degrees of freedom. Then the $β_{PSCL}$ parameter gives a complete description. This is also supported by the strong correlations $r_{PSCL} ≈ 0.80$ for these series.

For the three other series: spring, summer and fall, the DW-test gives positive or negative autocorrelations, indicating that the linear relation found is not a complete description. These series should be further analyzed for development of better models. The Durbin-Watson tables [26] show lower and upper limits (D(L) and D(U)) based on number of observations and number of parameters excluding the intercept in the regression model.

**TABLE 3.**
**Durbin Watson test on the autocorrelations in the PSCL model residuals**

| Series | DW | level result | Result |
|---|---|---|---|
| Year | 2.10 | 4-D(U)>DW>D(U) | No autocorrelation |
| Winter | 2.44 | 4-D(U)>DW>D(U) | No autocorrelation |
| Spring | 0.56 | DW<D(L) | Positive autocorrelation |
| Summer | 0.64 | DW<D(L) | Positive autocorrelation |
| Fall | 3.24 | DW > 4-D(L) | Negative autocorrelation |

*The number of observations in each series is 9. Because of inspection of the data on beforehand one degree of freedom has been subtracted which corresponds to 8 instead of 9 "effective" observations. The model has one parameter in addition to the intercept. Then the 5% significance levels for the DW test are D(L)= 0.763 and D(U)=1.332. The DW test is considered to have no significant autocorrelations if D(U)<DW< 4-D(U), indifferent if D(L)<DW<D(U) and 4-D(L)>DW > 4-D(U), positive autocorrelations if DW<D(L) and negative autocorrelations if DW>4-D(L).*

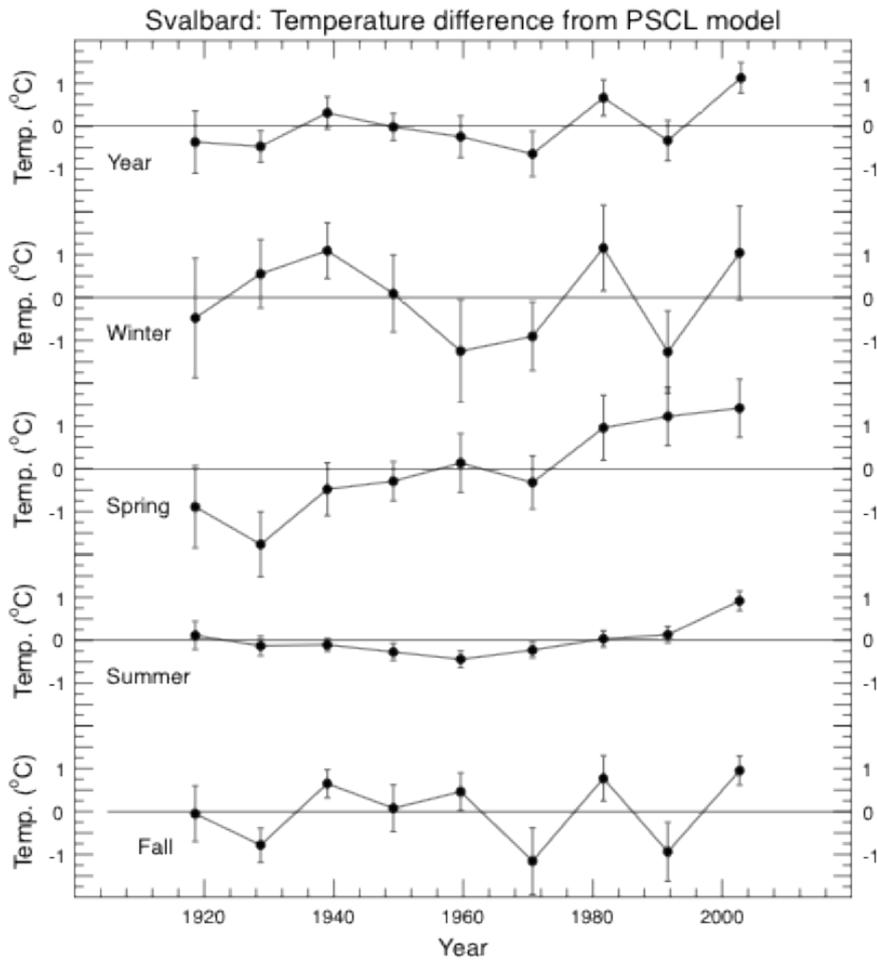

*Figure 5: Residuals from PSCL-model for the Svalbard temperature series and the seasonal series.*

The residuals from the PSCL-model are shown in figure 5. For the winter temperatures the trend in the residuals $\beta_2$=-0.5°C/100 yr. The spring residuals show $\beta_2$=3.2 °C/100 yr, which is a small reduction from $\beta_1$(table 1). For the year mean residuals we get $\beta_2$=1.1°C/100 yr

*2.5 Predictions for mean temperatures in sunspot cycle 24*
Of the 5 series investigated, the yearly mean and the winter mean temperatures are completely described by the PSCL-model. This model can be used to give forecasts for SC24 based on SCL23. The resulting forecasts show that the mean yearly temperature will decrease from -4.2 °C in SC23 to -7.8 °C, with a 95% confidence interval [-5.8 : -9.6]°C in SC24. This is the same result as with un-weighted relations in SSH11[15]. For the winter temperature the forecasts show a decrease from -11.2 in SC23 to -17.2 °C with a 95% confidence interval [-14 : -20.5]. These predictions are shown as diamonds in figure 3. For the other series: spring, summer and fall, the DW test and the less significance of the $\beta_{PSCL}$ values, make less confident predictions. They are shown as diamonds in figure 4.

## 3. Discussion

Our main result is a strong correlation between the mean air temperature at Svalbard in a solar cycle and the length of the previous solar cycle. The relation is highly significant for the yearly and winter mean temperatures. This is documented by stringent statistical tests showing no significant autocorrelations in the residuals, and small standard deviations in the $β_{PSCL}$-values, giving relations significant on the 95% level. For the spring and fall series the $β_{PSCL}$-values are significant at about 80% level, depending on how the autocorrelations reduce the degrees of freedom. For the summer temperatures there is no significant PSCL-relation. The differences through the seasons may partly be explained by local conditions.

The yearly and winter solar influence on the Svalbard temperature is estimated to ≈60%. The fact that there appears to be a clear solar influence on the air temperature in Svalbard during the winter may come as a surprise, as the Sun at 78°13'N is below the horizon from October 28 to February 14. Consequently, there is very little incoming solar radiation during the period December-February. Most likely, the explanation should be sought in the recurrent advection of warm air masses from lower latitudes across the North Atlantic towards Svalbard. This usually happens several times each winter, and often results in marked temperature increases within few hours. In this way a solar temperature signal originating at lower latitudes may be recorded at Svalbard, even during the winter.

In the spring (MAM) the landscape at Svalbard is almost completely snow covered, which means that most of the incoming short wave radiation from the Sun will be reflected, which results in little direct solar warming in the spring. On the other hand spring is normally the driest period of the year, and dominated by an Arctic anticyclone, which prohibits warm air advection from lower latitudes. Here is a marked difference between winter (DJF) and spring (MAM). The spring relation is modulated by ice or no-ice in the fjord. In the summer, fall and winter the fjord has been ice-free since 1912, but ice usually appears some times in the spring months (MAM). The increasing spring temperatures may be related to less ice in the fjord, and the general reduction of the Arctic ice.

The summer air temperature recorded at Svalbard Airport is highly influenced by local wind conditions, partly controlled by a land-sea breeze effects because of relatively small regional air pressure differences during the summer. By this the summer air temperature is controlled mainly by local conditions in the neighbourhood of the meteorological station (topography, land surface characteristics and surface temperature in the adjoining fjord), which are relatively stable from summer to summer.

Looking at the observed and averaged temperatures in figure 3, we get the impression of a periodic variation with a period of about 70 years. This may be the Low Frequency Oscillation in the Arctic temperatures as determined by Polyakov et al. [22]. However, if we study the residuals in figure 5, these oscillations have more or less disappeared. We may then speculate if they are related to solar variability described by the PSCL-model. The remaining trend for the MAAT: $β_2$=1.1 °C/100 yrs agrees with the trend 0.9±0.3 °C/100 yrs, determined by Polyakov et al. [22] for the Arctic poleward of 62N.

The strong trend in the spring residuals of 3.2 °C/100 yrs, may be explained as a Polar amplification as described by Bekryaev et al. [27], who find stronger trends in Polar spring warming than for the other seasons. For the period 1901-2008, they find for the Northern Polar Area trends for the annual, winter and fall temperatures, in agreement with our results ($β_2$). For the spring and summer they find trends of 1.6 and 0.88 °C/100 years, which is in agreement with our summer residuals trend, but about one half of our spring trend of 3.2 °C/100 yrs. They find an increasing trend (Polar amplification) when they determine trends for recent periods 1959-2008 and 1979-2008, the latter with an annual trend of 6.4 °C/100 yrs [27]. We cannot confirm this result in our residual series, but agree to a possible explanation of the stronger amplification of the spring trends by an albedo-ice effect.

The result of our PSCL-model, explaining more than 60% of the temperature variance for annual and winter temperatures for Svalbard, can be compared with a solar forcing well over 75% of the variance for the decadally-smoothed Arctic mean or spring temperatures as determined by Soon [28]. His analysis is based on decadal and multi-decadal reconstruction of Arctic, North of 62N, temperatures from wavelet analysis and correlation of the reconstructed temperatures with solar TSI variations.

The lag of one solar cycle for the temperature response may have two explanations. The first is a relation between solar cycle length and the amplitude ($R_{max}$) of the next cycle as found by Solanki and Krivova [11], assuming a relation between $R_{max}$ and temperature. This effect should be global. The other factor is the transport time of a solar signal with the Atlantic currents from the warm Caribbean to the West Coast of Svalbard. Analyzing sea-temperatures in the Faroe-Shetland Channel and the Kola Section, Yndestad et al. [29] find a phase delay of 2 years for a lunar-tide signal. A solar irradiation signal from the Caribbean may therefore take several years to reach the Svalbard region. Increased sea temperature will also reinforce the advective fall and winder warm air flows, and explain seasonal differences at Svalbard. Additional arguments for this interpretation is the higher PSCL-correlations found at the North Atlantic stations (Torshavn, Akureyri, Svalbard) than in costal and inland regions of Norway by SSH11 [15].

Based on the PSCL-relation we predict a temperature decrease at Svalbard of about ≈3.5 °C [±2 °C - 95% confidence interval] in the period 2009-20 compared with the previous SC23. This drop is of the same order as a forecast based on the strongest temperature cycles in HSS12 [3]. For the average winter temperatures a temperature drop of ≈6 °C is estimated and shown in figure 3 as a diamond with bars indicating the 95% confidence interval. In this figure also observed temperatures 2009-2011 are shown as broken lines, giving an impression that at least the winter temperature has decreased substantially already.

## 4. Conclusions

- A linear relation exists in the temperature series from Svalbard between the length of a solar cycle and the average temperature in the next solar cycle.

- The yearly average and the winter temperatures can be modelled as a function of the length of the previous solar cycle, with highly significant negative trends. We call this the PSCL regression model.

-The residuals from the PSCL-model show no positive autocorrelation using the Durbin Watson test. The estimated correlation coefficients between the observed temperatures and the temperature from the fit to the regression models are reasonably high for yearly average and winter temperatures. Also the uncertainty levels of the estimated correlations coefficients calculated by Bootstrap analysis are on an acceptable level. Hence, the winter model and the yearly average model are considered to be acceptable, which means that no additional variables are needed.

- A measure of the solar contribution is the coefficient of determination $r^2 \approx 0.6$ for the PSCL-year and winter models. This indicates that about 60% of the temperature variation can be attributed to solar activity for the yearly average and the winter average temperatures.

-For the average winter temperature the residuals shows a negative linear trend, which indicates that cooling might have taken place the last 100 years if the solar activity did not increase as observed by the shortening of the solar cycle.

-The solar cycle/temperature relation (our Model PSCL) can, when a sunspot cycle is finished, be used to predict the temperature in the next solar cycle. For Svalbard it means an estimated cooling of $\approx 3$ °C for the yearly average temperature from SC23 to the present SC24, which will last at least until 2020. The winter cooling will be $\approx 6$ °C. These predictions can test a possible solar-climate connection.

- This regression forecasting model benefits, as opposed to the majority of other regression models with explanatory variables, to use an explanatory variable – the previous sun cycle length – nearly without uncertainty. Usually the explanatory variables have to be forecasted, which of cause induce significant additional forecasting uncertainties.

- The negative trends in the spring and fall PSCL models are significant on 80% level. With positive and negative autocorrelations in the residuals, one may expect also other variables could be present for these series. The spring model residuals show a significant secular trend of 0.032 °C/yr, which indicates an amplification of some kind, probably related to diminishing Arctic and local ice cover in the spring season with an albedo effect. The residuals from the fall series show no significant trend. This may be explained by the nearby fjord (Isfjord) which never has been frozen in the fall and winter seasons since the start of the temperature series in 1912.

**Acknowledgements**
The Svalbard temperature data series used in this study was obtained from the *eKlima* internet data portal run by the *Norwegian Meteorological Institute*. Håkon Stordahl is thanked for help with the Bootstrap programming, and an anonymous referee for suggestions that helped us to improve the paper considerably.

15